\documentclass[11pt,a4paper]{article}
\usepackage[margin=24mm]{geometry}
\usepackage[T1]{fontenc}
\usepackage{lmodern}
\usepackage{amsmath,amssymb,amsthm,mathtools,booktabs,array,longtable,microtype}
\usepackage[colorlinks=true,linkcolor=blue,citecolor=blue,urlcolor=blue]{hyperref}
\hypersetup{pdftitle={Low-Weight Canonical Logical Bases from Pair-Partition Codes},pdfauthor={Koki Okada and Kenta Kasai}}
\newcommand{\row}{\operatorname{row}}
\newcommand{\F}{\mathbb F}\newcommand{\rank}{\operatorname{rank}}\newcommand{\adj}{\operatorname{adj}}

\makeatletter
\newenvironment{pairedmatrices}[1]{%
  \begingroup
  \renewcommand*\env@matrix{\hskip -\arraycolsep
    \let\@ifnextchar\new@ifnextchar
    \array{*\c@MaxMatrixCols{w{c}{#1}}}}%
}{\endgroup}
\makeatother

\title{Low-Weight Canonical Logical Bases\\from Pair-Partition Codes}
\author{Koki Okada \qquad Kenta Kasai\\[3pt]\small Institute of Science Tokyo}
\date{}
\begin{document}\maketitle
\begin{abstract}
We construct complete canonical logical bases for qubit CSS codes by assigning quaternary coefficients to binary circulant permutation matrix pair-partition (CPM--PP) checks, constructing and normalizing logical representatives, and expanding the result into binary matrices. When the two check systems have invertible submatrices on disjoint column sets, cofactor representatives are canonically paired by the inverse of a single pairing polynomial. The resulting pairs span the entire logical space. When the pairing polynomial is a cyclic-shift monomial with coefficient one, normalization preserves the binary weights of the representatives. We state the construction for general block dimensions and CPM size, work through a corresponding example, and report the parameters, check ranks, and basis weights of seven binary codes. Representative examples have parameters $[[320,80,14]]$, $[[448,112,18]]$, and $[[2048,512,24]]$. All three have maximum check weight 10 on both the X and Z sides. Their canonical logical representatives have binary weights 25, 27, and 27, respectively, on both the X and Z sides.
\end{abstract}

\section{Introduction}\label{sec:intro}
Let $H_X\in\F_2^{m_X\times n}$ and $H_Z\in\F_2^{m_Z\times n}$ be the binary check matrices of a CSS code~\cite{css,steane}, satisfying $H_XH_Z^T=0$. Here $n,m_X,m_Z\in\mathbb Z_{>0}$, and the number of logical qubits is $k=n-\rank H_X-\rank H_Z$. Our objective is to construct $k$ row representatives on each side, arranged as $L_X,L_Z\in\F_2^{k\times n}$, such that
\begin{equation}\label{eq:known-canonical}
 H_ZL_X^T=0,\qquad
 H_XL_Z^T=0,\qquad
 L_XL_Z^T=I_k
\end{equation}
The first two identities ensure that the representatives commute with the checks and preserve the code space. The last means that matching logical X and Z representatives anticommute, while those with different indices commute. Thus the rows specify physical X and Z operators for each encoded qubit. A basis is \emph{complete} when it contains all $k$ independent logical pairs. Such pairs can be obtained by linear algebra from the standard form of a stabilizer code~\cite[Sec.~4.1]{gottesman}. We give a polynomial construction that retains the cyclic structure of the checks.

Quasi-cyclic quantum LDPC constructions use circulant permutation matrices (CPMs) to specify paired sparse checks with controlled block dimensions~\cite{hi}. A subsequent construction assigns nonbinary coefficients to binary checks and replaces the resulting field elements by binary multiplication matrices~\cite[Secs.~II-B and II-C]{kasai}. The CPM--PP construction specifies commuting checks through pair partitions and matching exponent differences~\cite[Secs.~II-B and II-C]{pp}. These methods determine the check matrices. For logical representatives, the standard-form method above gives canonical pairs, while cofactor constructions for quasi-cyclic codes provide polynomial kernel vectors~\cite[Lemma~6]{sv}.

The framework in~\cite[Secs.~II and VII]{design} treats the choice of representatives for all logical degrees of freedom, and their X/Z pairing, as design tasks alongside the construction of check matrices. Cyclic shifts of short seeds need not span the full logical space~\cite[Sec.~VII.3]{design}. This motivates a construction that establishes the kernel conditions, canonical pairing, and completeness together.

We give such a construction when the two check systems admit \emph{invertible submatrices on disjoint column sets}. Cofactor seeds then have a common scalar polynomial pairing, whose inverse normalizes all logical pairs. The invertible submatrices determine the check ranks, so the number of pairs can be shown to equal the full logical dimension. We apply this condition to quaternary coefficients on CPM--PP checks and retain the polynomial description through binary expansion. The resulting explicit bases realize the completeness and canonical-pairing objectives in~\cite{design}.

The coefficient assignment is essential to the invertibility condition for a fully populated binary CPM array with $J>1$. Evaluating its monomial entries at $z=1$ gives an all-one matrix, so every $J\times J$ minor has determinant zero at $z=1$ and cannot be a unit. This obstruction is related to the trivial-packet rank deficiency discussed in~\cite[Sec.~VII.3]{design}. Quaternary coefficients can remove it while preserving commutation. The final binary code has length $2LP$; it is a new code constructed from the original length-$LP$ PP checks. Weight preservation during canonicalization is guaranteed here by the additional pure-shift condition on $g$ in~\eqref{eq:flow-shift-normalization}, which all seven examples satisfy. The tradeoff is an \emph{increase in binary check weight}: in the worked example, row weight rises from 8 in the initial binary CPM--PP checks to 10 after quaternary coefficient assignment and companion-matrix expansion. Ease of basis construction must therefore be considered together with the number of qubits involved in each check.

Section~\ref{sec:construction-flow} gives the four-stage construction for general $J,L,P$ and outlines its proof. Section~\ref{sec:examples} applies the stages with $J=3,L=8,P=20$ and reports seven code instances and their canonical logical bases. The coefficient field is fixed to $\F_4$.

\section{From CPM--PP checks to binary canonical logical bases}\label{sec:construction-flow}
The starting point is the three conditions for the binary logical matrices $L_X,L_Z\in\F_2^{k\times n}$:
\[
 H_ZL_X^T=0,\qquad H_XL_Z^T=0,\qquad L_XL_Z^T=I_k
\]
The first requires every logical X to commute with every Z check; the second requires every logical Z to commute with every X check. These conditions preserve the code space. The third pairs the logical operators: matching X and Z rows anticommute, whereas rows with different indices commute. Taking $k=n-\rank H_X-\rank H_Z$ pairs gives a complete logical basis. We seek such a basis with few nonzero entries in each row.

The check matrices $H_X,H_Z$ specify the code space; the logical matrices $L_X,L_Z$ specify physical representatives of X and Z for each encoded qubit. We first construct commuting CPM--PP checks and assign quaternary coefficients (Stages~1--2), then solve the three conditions over $\F_4$ (Stage~3) and expand into binary matrices (Stage~4). The last expansion preserves all three conditions. We give the procedure for general $J,L,P$; Section~\ref{sec:examples} follows the same stages for a concrete instance.

We follow the CPM--PP notation in~\cite[Sec.~III.1, Definition~1]{design}. For simplicity, set $J=J_X=J_Z$. Here $J$ is the number of block rows in each check matrix, $L$ is their common number of block columns, and $P$ is the CPM size. Let $J,P\in\mathbb Z_{>0}$ and $L\in2\mathbb Z_{>0}$, with $L>2J$ for the logical-basis construction, where $\mathbb Z_{>0}=\{1,2,\ldots\}$. Indices start at zero: $i,j\in\{0,\ldots,J-1\}$ and $\ell\in\{0,\ldots,L-1\}$. The initial binary checks are denoted by $H_X^{(0)},H_Z^{(0)}$, quaternary matrices carry the superscript $(4)$, and the final binary matrices are written as $H_X,H_Z,L_X,L_Z$ without a superscript.

\subsection{Stage 1: Construct binary CPM--PP check matrices}
For $s\in\mathbb Z_P$, let $C(s)\in\F_2^{P\times P}$ be the circulant permutation matrix whose nonzero entry in row $a\in\mathbb Z_P$ lies in column $a-s\pmod P$. We use the representatives $0,\ldots,P-1$ for the residue classes in $\mathbb Z_P$. Thus $C(s)$ sends position $t\in\mathbb Z_P$ of a column vector to position $t+s$. Choose a $J\times J$ array $M=(M_{ij})$, where each $M_{ij}$ partitions the block-column set $\{0,\ldots,L-1\}$ into $L/2$ disjoint pairs.

Choose exponent arrays $E=(e_{i\ell}),D=(d_{j\ell})\in\mathbb Z_P^{J\times L}$ satisfying, for every $0\le i,j<J$ and every $\{u,v\}\in M_{ij}$,
\begin{equation}\label{eq:flow-pp-shifts}
 d_{j,u}-e_{i,u}\equiv d_{j,v}-e_{i,v}\pmod P
\end{equation}
The entries $e_{i\ell},d_{j\ell}\in\mathbb Z_P$ specify CPM shifts; they are distinct from the quaternary coefficients introduced below. Set
\begin{equation}\label{eq:flow-plain}
 H_X^{(0)}=\bigl(C(e_{i\ell})\bigr)_{i,\ell},\qquad
 H_Z^{(0)}=\bigl(C(d_{j\ell})\bigr)_{j,\ell}
\end{equation}
so that $H_X^{(0)},H_Z^{(0)}\in\F_2^{JP\times LP}$. The $(i,j)$ block of their product is
\[
 \bigl(H_X^{(0)}(H_Z^{(0)})^T\bigr)_{ij}
 =\sum_{\ell=0}^{L-1}C(e_{i\ell}-d_{j\ell})=0
\]
because the two terms associated with each pair agree by~\eqref{eq:flow-pp-shifts} and cancel in characteristic two. This is the cancellation principle of the pair-partition (PP) construction~\cite[Sec.~III.1]{design}\cite[Secs.~II-B and II-C]{pp}. It does not require $P$ to be prime.

\subsection{Stage 2: Assign quaternary coefficients preserving PP cancellation}
Write $\F_4=\{0,1,\omega,\omega^2\}$ with $\omega^2=\omega+1$. Its nonzero elements are $1,\omega,\omega^2$, and arithmetic uses $a+a=0$ and $\omega^3=1$. In addition to the exponents $e_{i\ell},d_{j\ell}\in\mathbb Z_P$, introduce nonzero coefficients $\epsilon_{i\ell}\in\F_4^\times$ on the X side and $\delta_{j\ell}\in\F_4^\times$ on the Z side, where $\F_4^\times=\F_4\setminus\{0\}$. Define the blocks by
\begin{equation}\label{eq:flow-field}
 (H_X^{(4)})_{i\ell}=\epsilon_{i\ell}C(e_{i\ell}),\qquad
 (H_Z^{(4)})_{j\ell}=\delta_{j\ell}C(d_{j\ell})
\end{equation}
Each block belongs to $\F_4^{P\times P}$, and the full matrices satisfy $H_X^{(4)},H_Z^{(4)}\in\F_4^{JP\times LP}$. This replaces each nonzero entry of a CPM by its assigned quaternary coefficient.

For every pair $\{u,v\}\in M_{ij}$ from Stage~1, choose the coefficients to satisfy
\begin{equation}\label{eq:flow-pp-coeff}
 \epsilon_{i,u}\delta_{j,u}=\epsilon_{i,v}\delta_{j,v}
\end{equation}
Then
\[
 \bigl(H_X^{(4)}(H_Z^{(4)})^T\bigr)_{ij}
 =\sum_{\ell=0}^{L-1}\epsilon_{i\ell}\delta_{j\ell}C(e_{i\ell}-d_{j\ell})=0
\]
since the paired terms have both the same shift and the same coefficient, and hence cancel over $\F_4$. This coefficient condition ensures CSS commutation. The invertibility of the selected submatrices, needed below, is checked separately.

\subsection{Stage 3: Construct canonical logical bases in three steps}
Stage~2 has fixed the checks $H_X^{(4)},H_Z^{(4)}\in\F_4^{JP\times LP}$. We now seek $L_X^{(4)},L_Z^{(4)}\in\F_4^{sP\times LP}$, where $s=L-2J$, satisfying the same three conditions over $\F_4$:
\begin{equation}\label{eq:flow-canonical}
 H_Z^{(4)}(L_X^{(4)})^T=0,\qquad
 H_X^{(4)}(L_Z^{(4)})^T=0,\qquad
 L_X^{(4)}(L_Z^{(4)})^T=I_{sP}
\end{equation}
We satisfy the first two conditions by constructing seed vectors in the appropriate check kernels, and the third by normalizing their pairings. We then expand the seeds into all rows of the logical matrices. To retain the cyclic structure, calculate in
\begin{equation}\label{eq:flow-ring}
 R_P=\F_4[z,z^{-1}]/(z^P-1)
\end{equation}
Here coefficients lie in $\F_4$ and exponents are reduced modulo $P$. The ring $R_P$ need not be a field: all inverses below are required to exist in this ring, including when $P$ is even. The map $\Phi_P:R_P\to\F_4^{P\times P}$ replaces each polynomial by its circulant matrix:
\begin{equation}\label{eq:flow-Phi}
 \Phi_P\!\left(\sum_{a=0}^{P-1}c_a z^a\right)
 =\sum_{a=0}^{P-1}c_a C(a)
\end{equation}
where $c_a\in\F_4$. Thus the quaternary checks are the entrywise expansions of
\begin{equation}\label{eq:flow-polynomial-checks}
 \widehat H_X=(\epsilon_{i\ell}z^{e_{i\ell}}),\qquad
 \widehat H_Z=(\delta_{j\ell}z^{d_{j\ell}})
\end{equation}
with $\widehat H_X,\widehat H_Z\in R_P^{J\times L}$. A hat denotes a compact polynomial representation of a check matrix: each scalar monomial stands for an entire $P\times P$ block. These checks are given; the seed vectors constructed below are the unknowns. For $p\in R_P$, write $p^*(z)=p(z^{-1})$; for $U\in R_P^{m\times n}$, write $U^\dagger=(U^*)^T\in R_P^{n\times m}$, where $m,n\in\mathbb Z_{>0}$. Coefficients remain unchanged. This is the polynomial form of transposition: $\Phi_P(U^\dagger)=\Phi_P(U)^T$.

\paragraph{Step 3.1: Construct the seed vectors.}
Partition the block columns into disjoint ordered sets $A,B,I\subseteq\{0,\ldots,L-1\}$ with
\[
 |A|=|B|=J,\qquad |I|=s=L-2J\in\mathbb Z_{>0}
\]
A free column is a position whose seed entry is chosen first. The entries in $B$ adjust the X seed to satisfy the Z checks, and those in $A$ adjust the Z seed to satisfy the X checks. The X seed is zero on $A$, and the Z seed is zero on $B$. Define
\begin{equation}\label{eq:flow-minors}
 S_X=\widehat H_X[:,A],\quad S_Z=\widehat H_Z[:,B],\qquad
 \Delta_X=\det S_X,\quad\Delta_Z=\det S_Z
\end{equation}
where $[:,A]$ selects the columns in $A$ in order, $S_X,S_Z\in R_P^{J\times J}$, and $\Delta_X,\Delta_Z\in R_P$. For each free column $a\in I$, set all components of $\widetilde x_a,\widetilde z_a\in R_P^L$ to zero, then fill
\begin{equation}\label{eq:flow-seeds}
 \begin{aligned}
 (\widetilde x_a)_a&=\Delta_Z,&
 (\widetilde x_a)_B&=-\adj(S_Z)\widehat H_Z[:,a],\\
 (\widetilde z_a)_a&=\Delta_X,&
 (\widetilde z_a)_A&=-\adj(S_X)\widehat H_X[:,a]
 \end{aligned}
\end{equation}
The seeds $\widetilde x_a,\widetilde z_a\in R_P^L$ describe operator candidates, not quantum states. A term $c z^u$ in component $\ell$, with $c\in\F_4$ and $u\in\mathbb Z_P$, places coefficient $c$ at position $u$ in block $\ell$ of the unshifted quaternary candidate. Polynomial multiplication implements the cyclic shifts in the checks.

Here subscripts $A,B$ select the corresponding components, and $\adj(S)\in R_P^{J\times J}$ is the transpose of the cofactor matrix of $S\in R_P^{J\times J}$. It satisfies $S\adj(S)=\det(S)I_J$, which ensures $\widehat H_Z\widetilde x_a=\widehat H_X\widetilde z_a=0$. This is the cofactor kernel construction of~\cite[Lemma~6]{sv}, whose adjugate argument also holds over the commutative ring $R_P$. For example, component $b_r$ for $B=(b_0,\ldots,b_{J-1})$ is minus the determinant of $S_Z$ with column $r$ replaced by $\widehat H_Z[:,a]$. Minus and plus agree over $\F_4$.

This construction confines each seed to $J+1$ block columns. Each component is a determinant of a small monomial matrix, so for fixed small $J$ it consists of a small sum of monomials, with possible cancellations. Increasing $P$ changes the available positions without adding products to these determinant expressions. This produces sparse seeds; the next step determines whether canonicalization preserves their weight.

\phantomsection\label{step:flow-normalize}
\paragraph{Step 3.2: Normalize the pairings and preserve sparsity.}
The first two target conditions are already satisfied by the seeds. To obtain the third, examine their pairings. The X seed $\widetilde x_a\in R_P^L$ uses only $B\cup\{a\}$, whereas the Z seed $\widetilde z_a\in R_P^L$ uses only $A\cup\{a\}$. Since $A$ and $B$ are disjoint, every product involving an adjusting position vanishes. Only the common free position remains:
\[
 \widetilde x_a^\dagger\widetilde z_a
 =\sum_{\ell=0}^{L-1}(\widetilde x_a)_\ell^*(\widetilde z_a)_\ell
 =\Delta_Z^*\Delta_X
\]
For different free columns $a,b\in I$, the two supports are disjoint and the pairing is zero. Thus all same-index pairings are determined by one polynomial:
\begin{equation}\label{eq:flow-gram-poly}
 g(z):=\Delta_Z(z^{-1})\Delta_X(z)\pmod{z^P-1},\qquad
 \widetilde x_a^\dagger\widetilde z_b=
 \begin{cases}g(z)&a=b,\\0&a\ne b\end{cases}
\end{equation}
We call $g\in R_P$ the \emph{pairing polynomial}. It records the inner products between cyclic shifts of the seeds. The different-index pairings already have the desired value zero; only the common value $g$ must be changed to one.

Check that $g$ is a unit, meaning that $g^{-1}\in R_P$ exists. The units form $R_P^\times$, and the condition is equivalent to $\gcd(g,z^P-1)=1$. Compute the inverse by the extended Euclidean algorithm. If the condition fails, choose different column sets or revise the check coefficients or exponents. Nonzero $g$ alone is insufficient.

Keep each X seed and replace each Z seed by
\begin{equation}\label{eq:flow-normalized-seed}
 z_a'=\widetilde z_a g^{-1}\in R_P^L
\end{equation}

Then
\[
 \widetilde x_a^\dagger z_a'=gg^{-1}=1,\qquad
 \widetilde x_a^\dagger z_b'=0\quad(a\ne b)
\]

The Z-check condition for the unchanged X seeds is preserved, and the X-check condition for the new Z seeds is also preserved:
\[
 \widehat H_X z_a'=(\widehat H_X\widetilde z_a)g^{-1}=0
\]
Hence normalization gives the required pairings without violating either check condition. Under expansion, the polynomial one becomes $\Phi_P(1)=I_P$, so each seed pair gives $P$ canonically paired representatives, not just one.

Invertibility alone does not ensure that weights are preserved. An inverse with several terms combines shifted copies of a seed and may increase its weight. For example, $1+\omega z\in R_4$ is a unit because $(1+\omega z)^4=\omega^2$. Multiplication by this unit gives
\[
 \widetilde z_a(1+\omega z)=\widetilde z_a+\omega z\widetilde z_a
\]
By contrast, if $g=z^\nu$ with $\nu\in\mathbb Z_P$, its inverse is a pure shift, and
\begin{equation}\label{eq:flow-shift-normalization}
 z_a'=z^{-\nu}\widetilde z_a
\end{equation}
merely changes positions, leaving all coefficients unchanged. The quaternary weight is therefore preserved. After companion-matrix expansion, the same shift permutes pairs of binary coordinates, so the binary weight is also preserved by normalization. This is distinct from the possible weight increase caused by binary expansion itself. A general monomial $g=c z^\nu$ with $c\in\F_4^\times\setminus\{1\}$ additionally rescales the coefficients and need not preserve binary weight. All seven instances in Section~\ref{sec:examples} satisfy the stronger condition $g=z^\nu$ with coefficient one.

\paragraph{Step 3.3: Arrange the seeds as rows and expand.}
For $I=(a_0,\ldots,a_{s-1})$, define $\widehat L_X,\widehat L_Z\in R_P^{s\times L}$ by
\begin{equation}\label{eq:rp-logicals-general}
 (\widehat L_X)_{r,:}=\widetilde x_{a_r}^\dagger,\qquad
 (\widehat L_Z)_{r,:}=(g^*)^{-1}\widetilde z_{a_r}^\dagger
 \quad(0\le r<s)
\end{equation}
Here $(\widehat L_X)_{r,:}$ denotes row $r$, and likewise for $\widehat L_Z$. The reversal in $\dagger$ accounts for transposing column seeds into rows. Consequently, the Z column factor $g^{-1}$ becomes the row factor $(g^*)^{-1}$. Expand entrywise:
\begin{equation}\label{eq:flow-logical-expansion}
 L_X^{(4)}=\Phi_P(\widehat L_X),\qquad
 L_Z^{(4)}=\Phi_P(\widehat L_Z)
\end{equation}
Each polynomial row gives $P$ cyclically shifted rows, ordered by seed index $r$ and then shift $t=0,\ldots,P-1$. Thus $L_X^{(4)},L_Z^{(4)}\in\F_4^{sP\times LP}$ satisfy the three target conditions~\eqref{eq:flow-canonical}. The $sP$ pairs form complete logical bases; the proof outline below explains why.

For comparison, expanding the unnormalized seeds in the same order gives the \emph{pairing matrix}
\begin{equation}\label{eq:flow-gram}
 G=\operatorname{diag}(\underbrace{\Phi_P(g),\ldots,\Phi_P(g)}_{s\text{ copies}})
 \in\F_4^{sP\times sP}
\end{equation}
whose entries are the inner products of the unnormalized X and Z rows. Normalization changes this pairing to $I_{sP}$; computing and inverting the full matrix $G$ is unnecessary.

\subsection{Stage 4: Expand checks and logical bases into binary matrices}
Given $H_X^{(4)},H_Z^{(4)}\in\F_4^{JP\times LP}$ and $L_X^{(4)},L_Z^{(4)}\in\F_4^{sP\times LP}$, replace each entry $c\in\F_4$ by $\rho(c)\in\F_2^{2\times2}$. We use the companion-matrix representation of finite fields described in~\cite[Chap.~4, p.~106]{macwilliams}. Binary expansion of nonbinary CSS checks is used in~\cite[Sec.~II-C]{kasai}; here the symmetric multiplication matrices below allow the same representation on both sides. In the basis $(1,\omega)$, set
\begin{equation}\label{eq:flow-companion}
 \rho(0)=0_{2\times2},\quad \rho(1)=I_2,\quad
 \rho(\omega)=\begin{pmatrix}0&1\\1&1\end{pmatrix},\quad
 \rho(\omega^2)=\begin{pmatrix}1&1\\1&0\end{pmatrix}
\end{equation}
Here $\rho(\omega)$ is the companion matrix of $x^2+x+1$, and $\rho(\omega^2)=\rho(\omega)^2$. This is distinct from the CPM $C(s)$. Let $\mathcal R$ denote entrywise application of $\rho:\F_4\to\F_2^{2\times2}$, so that $U\in\F_4^{m\times n}$ gives $\mathcal R(U)\in\F_2^{2m\times2n}$. The final matrices are
\[
 H_X=\mathcal R(H_X^{(4)}),\qquad H_Z=\mathcal R(H_Z^{(4)}),
\]
\[
 L_X=\mathcal R(L_X^{(4)}),\qquad L_Z=\mathcal R(L_Z^{(4)})
\]
with dimensions
\begin{equation}\label{eq:flow-binary-dimensions}
 H_X,H_Z\in\F_2^{2JP\times2LP},\qquad
 L_X,L_Z\in\F_2^{2sP\times2LP}
\end{equation}

The map $\rho$ preserves addition and multiplication, and all four multiplication matrices are symmetric. Consequently, $\mathcal R(UV)=\mathcal R(U)\mathcal R(V)$ and $\mathcal R(U^T)=\mathcal R(U)^T$, giving
\begin{equation}\label{eq:flow-binary-check-identities}
 H_XH_Z^T=0,\qquad
 H_ZL_X^T=H_XL_Z^T=0,
\end{equation}
\begin{equation}\label{eq:flow-binary-canonical}
 L_XL_Z^T=I_{2sP}
\end{equation}
Under the unit condition in \hyperref[step:flow-normalize]{Step~3.2}, each check has binary rank $2JP$. The resulting code therefore has $n=2LP$ and $k=2(L-2J)P$.

\subsection{Proof outline: Validity and completeness}\label{sec:cofactor-readable}
The construction rests on four observations. Its hypotheses are CSS commutation, disjoint column sets $A,B,I$, and the unit condition on the pairing polynomial $g\in R_P^\times$. In particular, nonzero $g$ alone is not enough.

\paragraph{The representatives satisfy the checks.}
Substituting the cofactor entries in~\eqref{eq:flow-seeds} into the checks cancels the contribution from the free position by $S\adj(S)=\det(S)I$. Thus the X representatives belong to the Z kernel, and the Z representatives belong to the X kernel. Cyclic shifts and the linear combinations used for normalization preserve these kernel conditions.

\paragraph{The pairings are canonical.}
An X seed is supported in $B$ and its free position; a Z seed is supported in $A$ and its free position. Since $A,B,I$ are disjoint, only a shared free position contributes to the cross-pairing. Its contribution is the pairing polynomial $g=\Delta_Z^*\Delta_X\in R_P$ from~\eqref{eq:flow-gram-poly}. Multiplying the Z seeds by $g^{-1}\in R_P$ therefore makes the pairing matrix the identity.

\paragraph{The representatives span the full logical space.}
If $g=\Delta_Z^*\Delta_X$ is a unit, both factors are units, so $S_X,S_Z\in R_P^{J\times J}$ are invertible. Each expanded check consequently has quaternary rank $JP$, and the logical-space dimension is $LP-2JP=(L-2J)P=sP$. The relevant X space is the quotient $\ker H_Z^{(4)}/\row H_X^{(4)}$, identifying vectors that differ by an X stabilizer; the Z space is defined with X and Z exchanged. If a linear combination $cL_X^{(4)}$, with $c\in\F_4^{1\times sP}$, lies in $\row H_X^{(4)}$, pairing it with $(L_Z^{(4)})^T$ gives $c=0$: the canonical pairing~\eqref{eq:flow-canonical} gives $c$, whereas every X stabilizer pairs to zero. The same argument applies to Z. Thus the $sP$ classes on each side are independent and match the quotient dimension, proving completeness.

\paragraph{Binary expansion preserves the required identities.}
The multiplication matrices in~\eqref{eq:flow-companion} preserve addition and multiplication and are all symmetric. Binary expansion therefore respects products and transposes, carrying zero check pairings and identity logical pairings to their binary counterparts. The invertible submatrices also expand to invertible matrices, so the number of logical pairs is $2sP$, equal to the binary logical dimension.

Thus the disjoint invertible-submatrix condition provides a procedure from seed generation to complete binary canonical bases. This condition must be checked when selecting the columns and quaternary coefficients; it need not hold for every PP code.

\section{A worked construction and seven code instances}\label{sec:examples}
We apply Stages~1--4 of Section~\ref{sec:construction-flow} with $J=3,L=8,P=20$, and then report seven codes and their complete canonical logical bases obtained by the same procedure.

\subsection{Stage 1 example: Binary CPM--PP checks}
Choose the exponent arrays
\begin{pairedmatrices}{1.2em}
\begin{align}
\label{eq:flow-example-E}
E&=\begin{pmatrix}
 0&0&0&0&0&0&0&0\\
 0&2&16&12&13&18&10&9\\
 0&6&2&13&18&17&1&5
 \end{pmatrix},\\[6pt]
\label{eq:flow-example-D}
D&=\begin{pmatrix}
 0&9&18&17&18&0&17&9\\
 0&18&3&5&0&18&3&5\\
 0&1&17&13&13&17&1&0
 \end{pmatrix}
\end{align}
\end{pairedmatrices}
where $E,D\in\mathbb Z_{20}^{3\times8}$. The pair partitions can be specified explicitly. Writing $05$ for the unordered pair $\{0,5\}$, define
\[
 \begin{array}{ll}
 M_1=\{05,17,24,36\},&M_2=\{04,15,26,37\},\\
 M_3=\{07,16,25,34\},&M_4=\{04,15,23,67\},\\
 M_5=\{07,13,25,46\}
 \end{array}
\]
and use
\[
 M=\begin{pmatrix}M_1&M_2&M_3\\M_3&M_1&M_4\\M_2&M_5&M_1\end{pmatrix}
\]
For example, at $(i,j)=(0,0)$, the pair $\{1,7\}$ satisfies $d_{0,1}-e_{0,1}=9=d_{0,7}-e_{0,7}$. The other pairs also satisfy the PP shift condition~\eqref{eq:flow-pp-shifts}. Applying the binary check construction~\eqref{eq:flow-plain}, replace each exponent by the corresponding $20\times20$ CPM to obtain $H_X^{(0)},H_Z^{(0)}\in\F_2^{60\times160}$. For instance, block $(1,2)$ of $H_X^{(0)}$ is $C(16)$.

\subsection{Stage 2 example: Quaternary coefficients and polynomial checks}
For $J=3,L=8,P=20$, choose the coefficient arrays
\begin{pairedmatrices}{1.3em}
\begin{align}
\label{eq:flow-epsilon}
(\epsilon_{i\ell})&=\begin{pmatrix}
 1&\omega&\omega^2&1&\omega&1&1&\omega^2\\
 \omega&1&1&\omega^2&1&\omega&\omega^2&1\\
 \omega^2&1&1&\omega&1&\omega^2&\omega&1
 \end{pmatrix},\\[6pt]
\label{eq:flow-delta}
(\delta_{j\ell})&=\begin{pmatrix}
 1&\omega^2&\omega&1&\omega^2&1&1&\omega\\
 \omega&1&1&\omega^2&1&\omega&\omega^2&1\\
 \omega^2&1&1&\omega&1&\omega^2&\omega&1
 \end{pmatrix}
\end{align}
\end{pairedmatrices}
For $(i,j)=(0,0)$ and the pair $\{1,7\}$, the coefficient products agree: $\omega\omega^2=1$ and $\omega^2\omega=1$. Every other pair satisfies the coefficient condition~\eqref{eq:flow-pp-coeff} as well. By the weighted-block definition~\eqref{eq:flow-field}, block $(0,1)$ is $\omega I_{20}$ in $H_X^{(4)}$ and $\omega^2C(9)$ in $H_Z^{(4)}$. This gives $H_X^{(4)},H_Z^{(4)}\in\F_4^{60\times160}$.

\paragraph{Polynomial check matrices.}
Specialize the ring~\eqref{eq:flow-ring} to $P=20$. Substituting the arrays~\eqref{eq:flow-example-E}--\eqref{eq:flow-delta} into the polynomial check definition~\eqref{eq:flow-polynomial-checks} gives the following $3\times8$ matrices over $R_{20}=\F_4[z,z^{-1}]/(z^{20}-1)$:
\begin{pairedmatrices}{2.6em}
\begin{align}
\label{eq:rp-HX}
\widehat H_X&=\begin{pmatrix}
1&\omega&\omega^2&1&\omega&1&1&\omega^2\\
\omega&z^{2}&z^{16}&\omega^2z^{12}&z^{13}&\omega z^{18}&\omega^2z^{10}&z^{9}\\
\omega^2&z^{6}&z^{2}&\omega z^{13}&z^{18}&\omega^2z^{17}&\omega z&z^{5}
\end{pmatrix},\\[6pt]
\label{eq:rp-HZ}
\widehat H_Z&=\begin{pmatrix}
1&\omega^2z^{9}&\omega z^{18}&z^{17}&\omega^2z^{18}&1&z^{17}&\omega z^{9}\\
\omega&z^{18}&z^{3}&\omega^2z^{5}&1&\omega z^{18}&\omega^2z^{3}&z^{5}\\
\omega^2&z&z^{17}&\omega z^{13}&z^{13}&\omega^2z^{17}&\omega z&1
\end{pmatrix}
\end{align}
\end{pairedmatrices}

\subsection{Stage 3 example: Constructing and canonicalizing logical representatives}
\paragraph{Step 3.1: Constructing the seeds.}
Take $A=(0,1,2)$, $B=(4,5,7)$, and $I=(3,6)$. Then $s=8-2\cdot3=2$, and the column-selection rule~\eqref{eq:flow-minors} gives
\begin{pairedmatrices}{2.6em}
\[
\begin{aligned}
S_X&=\begin{pmatrix}
 1&\omega&\omega^2\\
 \omega&z^2&z^{16}\\
 \omega^2&z^6&z^2
 \end{pmatrix},\\[6pt]
S_Z&=\begin{pmatrix}
 \omega^2z^{18}&1&\omega z^9\\
 1&\omega z^{18}&z^5\\
 z^{13}&\omega^2z^{17}&1
 \end{pmatrix}
\end{aligned}
\]
\end{pairedmatrices}
Here $S_X,S_Z\in R_{20}^{3\times3}$, with arithmetic in $R_{20}=\F_4[z,z^{-1}]/(z^{20}-1)$. Their determinants $\Delta_X,\Delta_Z\in R_{20}$ are
\begin{equation}\label{eq:flow-example-dets}
 \Delta_X=z^4+z^6+z^{16},\qquad
 \Delta_Z=z^6+z^{16}+z^{18}
\end{equation}
For example, the six products contributing to $\Delta_X$ are $z^4,z^2,\omega^2z^2,z^{16},z^6,\omega z^2$. The three terms with exponent two cancel because $1+\omega+\omega^2=0$.

Using the seed formula~\eqref{eq:flow-seeds} at the free column $a=3\in I$, construct $\widetilde x_3,\widetilde z_3\in R_{20}^8$ using the columns in $R_{20}^3$ given by
\begin{pairedmatrices}{2.6em}
\[
\begin{aligned}
\widehat H_Z[:,3]&=\begin{pmatrix}z^{17}&\omega^2z^5&\omega z^{13}\end{pmatrix}^{T},\\[3pt]
\widehat H_X[:,3]&=\begin{pmatrix}1&\omega^2z^{12}&\omega z^{13}\end{pmatrix}^{T}
\end{aligned}
\]
\end{pairedmatrices}
Place $\Delta_Z$ in component 3 of $\widetilde x_3$ and fill components 4, 5, and 7 using the three column-replacement determinants. For example, component 4 is
\[
 (\widetilde x_3)_4=
 \det\begin{pmatrix}
 z^{17}&1&\omega z^9\\
 \omega^2z^5&\omega z^{18}&z^5\\
 \omega z^{13}&\omega^2z^{17}&1
 \end{pmatrix}
 =1+\omega^2z^5+\omega^2z^{11}+\omega z^{15}+\omega z^{18}+\omega^2z^{19}.
\]
The remaining components are computed in the same way. All components after canonicalization and conversion to the row representation are listed in~\eqref{eq:rp-LX}--\eqref{eq:rp-alpha-beta} and the polynomial lists below.

For the second free column $a=6\in I$, construct $\widetilde x_6,\widetilde z_6\in R_{20}^8$ using the two columns in $R_{20}^3$
\begin{pairedmatrices}{2.6em}
\[
\begin{aligned}
\widehat H_Z[:,6]&=\begin{pmatrix}z^{17}&\omega^2z^3&\omega z\end{pmatrix}^{T},\\[3pt]
\widehat H_X[:,6]&=\begin{pmatrix}1&\omega^2z^{10}&\omega z\end{pmatrix}^{T}
\end{aligned}
\]
\end{pairedmatrices}
Place $\Delta_Z$ and $\Delta_X$ in the respective components at position 6. Fill X positions 4, 5, and 7 and Z positions 0, 1, and 2 using~\eqref{eq:flow-seeds} or the column-replacement determinants. Repeating this calculation for every $a\in I$ is the seed-construction procedure for general $J,L$.

\paragraph{Step 3.2: Normalizing the pairings.}
For $a=3$, the X seed uses block columns $\{3,4,5,7\}$ and the Z seed uses $\{0,1,2,3\}$; only column 3 is shared. For $a=6$, only column 6 is shared. Seeds from different free columns have no common nonzero block position. Thus only the free entries contribute to same-index pairings, and different-index pairings vanish. Each free column contributes 20 shifted representatives, giving 40 candidates on each side.
Substituting the determinants~\eqref{eq:flow-example-dets} into the pairing-polynomial definition~\eqref{eq:flow-gram-poly} gives
\[
 g=(z^{-6}+z^{-16}+z^{-18})(z^4+z^6+z^{16})=z^6
 \quad(z^{20}=1)
\]
The block-diagonal pairing formula~\eqref{eq:flow-gram} then gives $G=\operatorname{diag}(C(6),C(6))$: within each group of 20 candidates, the partner index is shifted by six.

Since $g=z^6\in R_{20}^\times$, its inverse is $g^{-1}=z^{14}\in R_{20}$. The seed normalization~\eqref{eq:flow-normalized-seed} therefore takes the pure-shift form~\eqref{eq:flow-shift-normalization}: add 14 to every exponent of each Z seed and reduce modulo 20. For a matched seed pair, the normalized inner product is $z^6z^{14}=z^{20}=1$; different-index pairings remain zero. To illustrate the shift on a polynomial component,
\[
 z^{-6}(1+\omega z^5)=z^{14}+\omega z^{19}\quad\text{in }R_{20}
\]
The positions change from 0 and 5 to 14 and 19, but the two coefficients stay unchanged. The same holds for every component of the actual seeds. Consequently, the resulting $L_X^{(4)},L_Z^{(4)}\in\F_4^{40\times160}$ are canonically paired without increasing weight during normalization.
\paragraph{Step 3.3: Arranging the rows and expanding.}
Define $\alpha,\beta,p_0,\ldots,p_5,q_0,\ldots,q_5\in R_{20}$ below. Applying the canonical row formula~\eqref{eq:rp-logicals-general} with $I=(3,6)$ gives the two $2\times8$ matrices
\begin{pairedmatrices}{1.4em}
\begin{align}
\label{eq:rp-LX}
\widehat L_X&=\begin{pmatrix}
 0&0&0&\alpha&p_0&p_1&0&p_2\\
 0&0&0&0&p_3&p_4&\alpha&p_5
 \end{pmatrix},\\[6pt]
\label{eq:rp-LZ}
\widehat L_Z&=\begin{pmatrix}
 q_0&q_1&q_2&\beta&0&0&0&0\\
 q_3&q_4&q_5&0&0&0&\beta&0
 \end{pmatrix}
\end{align}
\end{pairedmatrices}
The free-position entries in the row formula~\eqref{eq:rp-logicals-general} are $\alpha=\Delta_Z^*$ and $\beta=(g^*)^{-1}\Delta_X^*$, giving
\begin{equation}\label{eq:rp-alpha-beta}
 \alpha=z^2+z^4+z^{14},\qquad \beta=1+z^2+z^{10}
\end{equation}
The remaining X entries are
\begin{align*}
 p_0&=1+\omega^2z+\omega z^{2}+\omega z^{5}+\omega^2z^{9}+\omega^2z^{15},\\
 p_1&=z^{3}+z^{4}+z^{5}+z^{13}+\omega z^{17}+\omega^2z^{18},\\
 p_2&=1+\omega^2z^{2}+\omega^2z^{6}+\omega z^{7}+\omega z^{11}+\omega z^{12},\\
 p_3&=\omega^2z+\omega z^{5}+\omega^2z^{11}+z^{12}+\omega z^{14}+\omega^2z^{17},\\
 p_4&=z^{3}+z^{5}+\omega^2z^{10}+z^{15}+z^{16}+\omega z^{19},\\
 p_5&=z^{2}+\omega z^{3}+\omega^2z^{4}+\omega^2z^{6}+\omega z^{12}+\omega z^{19}.
\end{align*}

The remaining Z entries are
\begin{align*}
 q_0&=z^{2}+z^{4}+\omega z^{8}+z^{11}+z^{12}+\omega^2z^{17},\\
 q_1&=\omega z^{4}+\omega^2z^{10}+\omega^2z^{12}+\omega z^{13}+z^{14}+\omega z^{17},\\
 q_2&=\omega+\omega^2z^{4}+\omega^2z^{8}+\omega z^{11}+z^{13}+\omega^2z^{14},\\
 q_3&=z^{2}+z^{3}+z^{4}+\omega^2z^{9}+\omega z^{10}+z^{14},\\
 q_4&=\omega z^{4}+\omega z^{5}+\omega z^{9}+\omega^2z^{10}+\omega^2z^{14}+z^{16},\\
 q_5&=\omega+\omega z^{3}+\omega^2z^{4}+z^{5}+\omega^2z^{10}+\omega^2z^{16}.
\end{align*}
Here $g=z^6$, so the Z row-normalization factor is $(g^*)^{-1}=z^6\in R_{20}$; it has already been incorporated into $q_i$ and $\beta$. All exponents are reduced modulo 20.

The unnormalized column seeds can also be written explicitly using these polynomials:
\begin{pairedmatrices}{2.7em}
\[
\begin{aligned}
\widetilde x_3&=\begin{pmatrix}0&0&0&\Delta_Z&p_0^*&p_1^*&0&p_2^*\end{pmatrix}^{T},\\[4pt]
\widetilde x_6&=\begin{pmatrix}0&0&0&0&p_3^*&p_4^*&\Delta_Z&p_5^*\end{pmatrix}^{T},\\[4pt]
\widetilde z_3&=\begin{pmatrix}z^6q_0^*&z^6q_1^*&z^6q_2^*&\Delta_X&0&0&0&0\end{pmatrix}^{T},\\[4pt]
\widetilde z_6&=\begin{pmatrix}z^6q_3^*&z^6q_4^*&z^6q_5^*&0&0&0&\Delta_X&0\end{pmatrix}^{T}
\end{aligned}
\]
\end{pairedmatrices}
All four vectors belong to $R_{20}^8$. Here $p^*(z)=p(z^{-1})$, and $\Delta_X,\Delta_Z$ are given in~\eqref{eq:flow-example-dets}. The factor $z^6$ in the Z seeds undoes the previous normalization after reversing exponents. These two seeds on each side determine all 40 candidate rows by the shift-and-transpose rule of Step~3.3.

The circulant map~\eqref{eq:flow-Phi}, applied to the checks and to the logical expansion~\eqref{eq:flow-logical-expansion}, gives $H_X^{(4)},H_Z^{(4)}\in\F_4^{60\times160}$ and $L_X^{(4)},L_Z^{(4)}\in\F_4^{40\times160}$. The subsequent companion-matrix expansion produces binary matrices of sizes $120\times320$ and $80\times320$, respectively.

\subsection{Stage 4 example: Binary companion-matrix expansion}
Substituting $J=3,L=8,P=20$ into the binary dimension formulas~\eqref{eq:flow-binary-dimensions} gives $H_X,H_Z\in\F_2^{120\times320}$ and $L_X,L_Z\in\F_2^{80\times320}$, with $n=320,k=80$. Using the companion matrices~\eqref{eq:flow-companion}, a quaternary check block $\omega C(e)$ becomes the $40\times40$ binary block $C(e)\otimes\rho(\omega)$. Here the Kronecker product $\otimes$ replaces each entry of the left matrix by that entry times the right matrix. Apply the same operation to the canonical logical bases. The initial binary checks have row weight $L=8$, and each row of the quaternary checks also has eight nonzero entries. However, $\rho(\omega)$ and $\rho(\omega^2)$ in~\eqref{eq:flow-companion} each contain a row of weight two. For the chosen coefficient arrays, every expanded check row has weight 10 on both sides. Binary expansion therefore increases the number of nonzero entries per check as well as the matrix dimensions.

The binary check identities~\eqref{eq:flow-binary-check-identities} and canonical pairing~\eqref{eq:flow-binary-canonical} hold for these matrices. Both check ranks are 120 and the logical dimension is 80. The 80 rows of each of $L_X,L_Z\in\F_2^{80\times320}$ form complete canonical logical bases. Every row on either side has binary weight 27.

\subsection{Seven codes and their complete canonical logical bases}\label{sec:seven-results}
All seven instances use $J=3,L=8$, the coefficient arrays~\eqref{eq:flow-epsilon} and~\eqref{eq:flow-delta}, and the column sets $A=(0,1,2)$, $B=(4,5,7)$, and $I=(3,6)$. Table~\ref{tab:seven-inputs} specifies the CPM size $P\in\mathbb Z_{>0}$, the exponent array $D\in\mathbb Z_P^{3\times8}$, and the row shifts $r=(r_0,r_1,r_2)\in\mathbb Z_P^3$ for each instance. The other exponent array $E\in\mathbb Z_P^{3\times8}$ is determined by the column permutation
\[
 (q(0),\ldots,q(7))=(5,7,4,6,2,0,3,1)
\]
and the entrywise formula
\begin{equation}\label{eq:seven-E}
 e_{i\ell}=-d_{i,q(\ell)}+r_i+d_{0\ell}\pmod P,
 \qquad 0\le i<3,\quad0\le\ell<8.
\end{equation}
The table determines the polynomial checks through~\eqref{eq:flow-polynomial-checks}, the seeds through~\eqref{eq:flow-seeds}, and the canonical logical rows through~\eqref{eq:rp-logicals-general}. The circulant expansion~\eqref{eq:flow-Phi} and companion matrices~\eqref{eq:flow-companion} then specify all four binary matrices of every instance. The first row gives the $[[320,80,13]]$ example worked out above.

\begin{table}[htbp]\centering\small
\renewcommand{\arraystretch}{1.15}
\begin{tabular}{@{}ccl@{}}\toprule
$P$ & $r$ & $D$ \\\midrule
20&$(0,18,17)$&$\begin{pmatrix}0&9&18&17&18&0&17&9\\0&18&3&5&0&18&3&5\\0&1&17&13&13&17&1&0\end{pmatrix}$\\[5pt]
20&$(0,1,7)$&$\begin{pmatrix}0&6&2&18&2&0&18&6\\0&1&10&11&0&1&10&11\\0&2&7&13&13&7&2&0\end{pmatrix}$\\[5pt]
22&$(0,1,4)$&$\begin{pmatrix}0&3&1&4&1&0&4&3\\0&1&6&17&0&1&6&17\\0&12&4&18&18&4&12&0\end{pmatrix}$\\[5pt]
22&$(0,1,19)$&$\begin{pmatrix}0&18&8&4&8&0&4&18\\0&1&6&17&0&1&6&17\\0&20&19&3&3&19&20&0\end{pmatrix}$\\[5pt]
26&$(0,8,12)$&$\begin{pmatrix}0&2&19&8&19&0&8&2\\0&8&11&21&0&8&11&21\\0&19&12&16&16&12&19&0\end{pmatrix}$\\[5pt]
28&$(0,4,19)$&$\begin{pmatrix}0&11&16&27&16&0&27&11\\0&4&7&21&0&4&7&21\\0&9&19&13&13&19&9&0\end{pmatrix}$\\[5pt]
128&$(0,105,11)$&$\begin{pmatrix}0&84&54&10&54&0&10&84\\0&105&92&63&0&105&92&63\\0&32&11&67&67&11&32&0\end{pmatrix}$\\[5pt]
\bottomrule\end{tabular}
\caption{Exponent data for the seven instances. The three rows of $D$ and the row shifts $r$ determine $E$ through~\eqref{eq:seven-E}. Entries are represented by $0,\ldots,P-1$.}\label{tab:seven-inputs}
\end{table}

Use the partitions $M_1,\ldots,M_5$ defined in Section~\ref{sec:examples}. The codes with $P\in\{20,26\}$ use the array $M$ of the worked example. Those with $P\in\{22,28,128\}$ use
\[
 M=\begin{pmatrix}M_1&M_2&M_3\\M_3&M_1&M_2\\M_2&M_3&M_1\end{pmatrix}
\]
All pairs satisfy the exponent condition~\eqref{eq:flow-pp-shifts} and the coefficient condition~\eqref{eq:flow-pp-coeff}.

Table~\ref{tab:seven-results} reports the binary codes and canonical logical bases. In every instance, the pairing polynomial~\eqref{eq:flow-gram-poly} is $g=z^a\in R_P^\times$, so normalization follows the pure-shift formula~\eqref{eq:flow-shift-normalization}. The checks satisfy $H_X,H_Z\in\F_2^{6P\times16P}$ and the logical matrices satisfy $L_X,L_Z\in\F_2^{4P\times16P}$. Direct computation confirms the binary check identities~\eqref{eq:flow-binary-check-identities} and canonical pairing~\eqref{eq:flow-binary-canonical}, together with the ranks:
\[
 \rank H_X=\rank H_Z=6P,\qquad
 H_ZL_X^T=H_XL_Z^T=0,\qquad L_XL_Z^T=I_{4P}
\]
Thus the $4P$ representatives on each side cover all logical degrees of freedom. Every check row has weight 10 on both sides in all seven instances.

\begin{table}[htbp]\centering\small
\setlength{\tabcolsep}{3pt}
\begin{tabular}{@{}cccccccc@{}}\toprule
$P$ & $[[n,k,d]]$ & \shortstack{Girth\\$\F_4/\F_2$} & \shortstack{Column\\weight} & \shortstack{Row\\weight} & \shortstack{Check\\rank} & \shortstack{Pairing\\polynomial $g$} & \shortstack{Basis weight\\(count)}\\\midrule
20&$[[320,80,13]]$&$6/4$&$3,4$&10&120&$z^{6}$&$27\ (80)$\\
20&$[[320,80,14]]$&$6/4$&$3,4$&10&120&$z^{8}$&$25\ (80)$\\
22&$[[352,88,15]]$&$6/4$&$3,4$&10&132&$z^{17}$&$27\ (88)$\\
22&$[[352,88,16]]$&$6/4$&$3,4$&10&132&$z^{16}$&$27\ (88)$\\
26&$[[416,104,17]]$&$6/4$&$3,4$&10&156&$z^{19}$&$25\ (52),\;27\ (52)$\\
28&$[[448,112,18]]$&$6/4$&$3,4$&10&168&$z^{18}$&$27\ (112)$\\
128&$[[2048,512,24]]$&$6/4$&$3,4$&10&768&$z^{122}$&$27\ (512)$\\
\bottomrule\end{tabular}
\caption{Binary code parameters and canonical logical bases, in the input order of Table~\ref{tab:seven-inputs}. Girth lists the Tanner-graph girths of the quaternary and final binary checks, respectively; each value applies to both X and Z. Column weight, row weight, and rank refer to each final binary check matrix. Each has $4P$ columns of weight 3 and $12P$ of weight 4, and all $6P$ rows have weight 10. Before binary expansion, each quaternary check has column weight 3 and row weight 8. Basis-weight counts are per side and agree on X and Z. Every listed distance $d$ is exact.}\label{tab:seven-results}
\end{table}

Each listed distance $d$ was certified by exhaustively excluding nontrivial logical operators of weight at most $d-1$ and independently checking an operator of weight $d$. We use the zero-syndrome search principle of~\cite[Sec.~V-A, Algorithm~2]{pp}, adapted to quaternary symbols with binary costs $0,1,1,2$ for $0,1,\omega,\omega^2$. Nontriviality is tested by the binary pairing with the complete opposite logical basis. Simultaneous cyclic translation preserves the checks and the binary cost. Translating a nonzero symbol in its first occupied block to position zero gives $8\cdot3=24$ roots, covering all nonzero words for these lift sizes. Every branch through cost $d-1$ was completed or validly pruned; a zero-syndrome stabilizer partial word can be removed from any extension to leave a smaller word with the same logical label. A weight-preserving coordinate permutation exchanges the X and Z check spaces, and was verified for each instance; hence $d_X=d_Z=d$ throughout. In particular, the $[[2048,512,24]]$ code has exact distance $d=24$.

For every listed instance, the supports of each matched X/Z pair intersect in exactly one qubit. Basis weight denotes the number of ones in a constructed binary representative; it is distinct from the minimum distance of the code. The reported weight distributions and support-intersection sizes are computational results for these seven inputs.

\begin{samepage}
For separate X- and Z-error decoding, group the two binary coordinates associated with each quaternary column into one symbol $a+b\omega\in\F_4$, with $a,b\in\F_2$, and group the corresponding syndrome pairs in the same basis. The binary syndrome equations then become the quaternary equations defined by $H_Z^{(4)}$ for X errors and $H_X^{(4)}$ for Z errors. For a memoryless physical error model, using the induced four-symbol error probabilities, nonbinary belief propagation (BP), specifically the sum-product algorithm, can operate directly on these quaternary Tanner graphs~\cite[Sec.~III]{kasai}. For all seven codes, each such decoding graph has girth 6, column weight 3, and row weight 8. This grouped decoding representation preserves the physical binary checks, whose row weight remains 10.
\par\end{samepage}

\section*{Declaration of generative AI use}
The authors used OpenAI Codex to assist with manuscript drafting, translation, and revision; the development and review of mathematical arguments; and the preparation of computational verification code and examination of numerical results. The authors reviewed the AI-assisted material and take full responsibility for the mathematical claims, computational results, references, and final content of the manuscript.


\begin{thebibliography}{9}\interlinepenalty=10000\small\setlength{\itemsep}{3pt}\setlength{\parskip}{0pt}
\bibitem{css} A. R. Calderbank and P. W. Shor, ``Good Quantum Error-Correcting Codes Exist,'' \emph{Physical Review A} \textbf{54}, 1098--1106 (1996). \href{https://doi.org/10.1103/PhysRevA.54.1098}{doi:10.1103/PhysRevA.54.1098}.
\bibitem{steane} A. M. Steane, ``Multiple Particle Interference and Quantum Error Correction,'' \emph{Proceedings of the Royal Society A} \textbf{452}, 2551--2577 (1996). \href{https://doi.org/10.1098/rspa.1996.0136}{doi:10.1098/rspa.1996.0136}.
\bibitem{gottesman} D. Gottesman, \emph{Stabilizer Codes and Quantum Error Correction}, Ph.D. thesis, California Institute of Technology, 1997, Sec.~4.1. \href{https://arxiv.org/abs/quant-ph/9705052}{arXiv:quant-ph/9705052}.

\bibitem{design} J. Y. Lee, K. Okada, N. Maskara, K. Kasai, and H. Zhou, ``Design Principles for Ultra-High-Rate Quantum Codes,'' \href{https://arxiv.org/abs/2609.30069v1}{arXiv:2609.30069v1} (2026).
\bibitem{sv} R. Smarandache and P. O. Vontobel, ``Quasi-Cyclic LDPC Codes: Influence of Proto- and Tanner-Graph Structure on Minimum Hamming Distance Upper Bounds,'' \emph{IEEE Transactions on Information Theory} \textbf{58}(2), 585--607 (2012). \href{https://doi.org/10.1109/TIT.2011.2173244}{doi:10.1109/TIT.2011.2173244}.
\bibitem{macwilliams} F. J. MacWilliams and N. J. A. Sloane, \emph{The Theory of Error-Correcting Codes}, North-Holland Mathematical Library, vol.~16. Amsterdam: North-Holland, 1977. \href{https://neilsloane.com/doc/ms77.html}{Author book page}.
\bibitem{pp} K. Okada and K. Kasai, ``Pair-Partition Constructions for CPM-Based Quantum LDPC Codes,'' \href{https://arxiv.org/abs/2607.14091v4}{arXiv:2607.14091v4} (2026).
\bibitem{hi} M. Hagiwara and H. Imai, ``Quantum Quasi-Cyclic LDPC Codes,'' in \emph{Proc. IEEE Int. Symp. Information Theory (ISIT)}, 2007, pp.~806--810. \href{https://doi.org/10.1109/ISIT.2007.4557323}{doi:10.1109/ISIT.2007.4557323}.
\bibitem{kasai} K. Kasai, M. Hagiwara, H. Imai, and K. Sakaniwa, ``Quantum Error Correction Beyond the Bounded Distance Decoding Limit,'' \emph{IEEE Transactions on Information Theory} \textbf{58}(2), 1223--1230 (2012). \href{https://doi.org/10.1109/TIT.2011.2167593}{doi:10.1109/TIT.2011.2167593}.
\end{thebibliography}
\end{document}